\documentclass[
aip,reprint,
amsmath,amssymb,superscriptaddress,longbibliography
]{revtex4-1}

\usepackage{hyperref}
\usepackage{graphicx}
\usepackage{mathtools}
\usepackage{hyperref}
\usepackage{color}
\usepackage{xcolor}
\usepackage{diagbox}
\usepackage{tabularx}
\usepackage{tabu}
\usepackage{verbatim}

\usepackage{tikz}

\newcommand{\upbar}{\tikz[overlay] \draw (0,1em)--(0,0em);}
\newcommand{\downbar}{\tikz[overlay] \draw (0,.5em)--(0,-1em);}

\usepackage[export]{adjustbox}

\draft 

\begin{document}

\title{Legendre-spectral Dyson equation solver with super-exponential convergence}

\author{Xinyang Dong}
\affiliation{Department of Physics, University of Michigan, Ann Arbor, MI 48109, USA}
\author{Dominika Zgid}
\affiliation{Department of Chemistry, University of Michigan, Ann Arbor, MI 48109, USA}
\affiliation{Department of Physics, University of Michigan, Ann Arbor, MI 48109, USA}
\author{Emanuel Gull}
\affiliation{Department of Physics, University of Michigan, Ann Arbor, MI 48109, USA}
\author{Hugo~U.~R.~Strand}
\email{hugo.strand@gmail.com}
\affiliation{Department of Physics, Chalmers University of Technology, SE-412 96 Gothenburg, Sweden}
\affiliation{Center for Computational Quantum Physics, The Flatiron Institute, New York, NY 10010, USA}

\date{\today} 

\begin{abstract}
%
%
Quantum many-body systems in thermal equilibrium can be described by the imaginary time Green's function formalism.
%
%
%
However, the treatment of large molecular or solid \textit{ab inito} problems with a fully realistic Hamiltonian in large basis sets is hampered by the storage of the Green's function and the precision of the solution of the Dyson equation.
%
%
We present a Legendre-spectral algorithm for solving the Dyson equation that addresses both of these issues.
By formulating the algorithm in Legendre coefficient space, our method inherits the known faster-than-exponential convergence of the Green's function's Legendre series expansion.
In this basis, the fast recursive method for Legendre polynomial convolution, enables us to develop a Dyson equation solver with quadratic scaling.
%
%
We present benchmarks of the algorithm by computing the dissociation energy of the helium dimer He$_2$ within dressed second-order perturbation theory. For this system, the application of the Legendre spectral algorithm allows us to achieve an energy accuracy of $10^{-9} E_h$ with only a few hundred expansion coefficients.
%
%
\end{abstract}

\maketitle
\makeatletter
\let\toc@pre\relax
\let\toc@post\relax
\makeatother


%


\section{Introduction}

%
%
The equilibrium properties of many-body quantum systems can be described by the finite temperature imaginary-time Green's function formalism \cite{Abrikosov:1975aa}, which is widely applicable to condensed matter physics, quantum chemistry, and material science.
Applications include numerical methods for low energy effective model Hamiltonians such as
lattice Monte Carlo \cite{Blankenbecler:1981aa},
dynamical mean field theory \cite{Georges:1996aa}
and its extensions \cite{Toschi:2007aa, Rubtsov:2008aa, Maier:2005aa},
and diagrammatic Monte Carlo \cite{Prokof:2007aa}.
\textit{Ab initio} calculations using
the random phase approximation \cite{doi:10.1021/ct5001268},
self-consistent second order perturbation theory \cite{PhysRevB.63.075112,Dahlen:2005aa, Phillips:2014aa, Phillips:2015aa, Kananenka:2016aa, Kananenka:2016ab, Rusakov:2016aa, Welden:2016aa, Iskakov:2019aa},
Hedin's $GW$ aproach \cite{Hedin:1965aa, Aryasetiawan:1998aa, Stan:2009aa, Kutepov:2009aa, vanSetten:2015aa, Maggio:2017aa, Grumet:2018aa, Kutepov:2016aa, Kutepov:2017aa},
and self energy embedding theory \cite{Kananenka:2015aa, Lan:2015aa, Zgid:2017aa, Lan:2017aa, Lan:2017ab, Tran:2018aa, Rusakov:2019aa}
can also be formulated in imaginary time.
%

While the finite temperature Green's function formalism is very successful in applications to model Hamiltonians, its applicability to quantum chemistry and materials science remains limited to simple molecular and periodic problems. This is due to the necessity of simultaneously describing both the core and valence orbitals, which results in an energy scale that is difficult to describe by a single imaginary time/frequency grid. A simple equidistant Matsubara grid would contain millions of points, thus making the storage and manipulation of the Green's functions computationally costly. In contrast, a grid with only a small number of equidistant points will result in a poorly converged energy or density matrix, making calculations with $\mu$Hartree precision challenging. Such precision is necessary in applications where the evaluation of interaction energies \cite{Taylor2016:aa, Chalasinski1994:aa, Chalasinski2000:aa, Chalasinski1988:aa}, energies of conformers \cite{Podeszwa2006:aa}, or energies of high-lying excited states \cite{Sharma2014:aa, Li2019:aa} is needed. Consequently, it is important to develop a compact representation that yield highly converged properties.

%
With the standard approach using equidistant Matsubara frequency \cite{Matsubara:1955aa} grids with finite frequency cut-off, the imaginary time Green's function only converges to the analytical result linearly in the number of Matsubara frequencies. Amending the representation with a low order high frequency expansion results in polynomial convergence \cite{Nils:2002aa, Comanac:2007aa, Dario:2016aa}.
In practice, this is problematic, since for systems with a wide range of energy scales, the number of coefficients is controlled by the largest energy scale \cite{Kananenka:2016ab}.
Alternatives such as uniform power meshes have had some success \cite{Ku:2002aa,Ku:2000aa}. However, the most compact representations are achieved using a set of (orthogonal) continuous basis functions directly in imaginary time, such as orthogonal polynomials \cite{Boehnke:2011fk,PhysRevB.98.075127} or numerical basis functions \cite{Shinaoka:2017aa,Chikano:2018aa,Chikano:2019aa,LiJia:2020aa,Kaltak:2019aa}. The convergence of such a representation is faster than exponential,\cite{Boehnke:2011fk,PhysRevB.98.075127} and asymptotically superior to any polynomially converging representation.

%
In all imaginary time methods a central step besides the solution of the impurity problem is the solution of the Dyson equation for the single particle Green's function \cite{Negele:1998aa, Fetter:2003aa, Atland:2006nx, Stefanucci:2013oq}.
%
%
In the Matsubara frequency representation \cite{Matsubara:1955aa} the Dyson equation is diagonal and can be readily solved. However, the solution is plagued by the polynomial convergence with respect to the number of frequency coefficients used.
In imaginary time the Dyson equation is a non-trivial integro-differential equation with a mixed boundary condition. Recently an algorithm for solving the Dyson equation in imaginary time using the Chebyshev polynomials has been presented  \cite{PhysRevB.98.075127}. This algorithm preserves the exponential convergence of the orthogonal polynomial expansion \cite{Boehnke:2011fk}. However, the central convolution step has a cubic scaling in the expansion order $N_L$, $\sim \mathcal{O}(N_L^3)$, which limits the applicability of the algorithm.

The development of compact representations and algorithms for solving the Dyson equation is an active field of research, see Tab.~\ref{table:method_v2} for an overview of the state-of-the-art methods. For a recent development see Ref. \onlinecite{Kaltak:2019aa} .

\begin{table*}[bth]
  \begin{tabularx}{0.82\textwidth}{|l|l|c|c|c|}
 \hline 
 Domain & Basis & Convergence & Compactness & Dyson scaling\\
    \hline \hline
    Matsubara frequency
    & Finite frequency cutoff & $\mathcal{O}(1)$ & Poor & $\mathcal{O}(N)$ \\
    & Tail correction, $p$th order \cite{Nils:2002aa, Comanac:2007aa, Dario:2016aa} & $\mathcal{O}(N^{-p})$ & Fair & \downbar \\
    & Spline grid \cite{Kananenka2016:aa} & --- & Good & \upbar \\
    \hline \hline
    Both frequency and time
    & Sparse sampling \cite{LiJia:2020aa} & --- & see Ref.~\onlinecite{LiJia:2020aa} \footnote{The compactness of the sparse sampling approach depends on the real-time basis employed.} & $\mathcal{O}(N)$ \\ & Minimax Isometry \cite{Kaltak:2019aa} & --- & see Ref.~\onlinecite{Kaltak:2019aa} & $\mathcal{O}(N)$ \\
    \hline \hline
    Imaginary time
    & Uniform mesh & $\mathcal{O}(N^{-1})$ & Poor & $\mathcal{O}(N^3)$ \\
    & Power mesh \cite{Ku:2002aa,Ku:2000aa,PhysRevB.97.115164} & --- & Fair & $\mathcal{O}(N^3)$ \\
    \hline \hline
    Orthogonal functions
    & Intermediate representation \cite{Shinaoka:2017aa,Chikano:2018aa,Chikano:2019aa} & $\lesssim \mathcal{O}(e^{-N})$ & Excellent & No \\
    & Chebyshev polynomials \cite{PhysRevB.98.075127} & $\lesssim \mathcal{O}(e^{-N})$ & Very Good & $\mathcal{O}(N^3)$ \\
    & Legendre polynomials (\textit{this work}) & $\lesssim \mathcal{O}(e^{-N})$ \cite{Boehnke:2011fk} & Very Good & $\mathcal{O}(N^2)$ \\
    \hline
  \end{tabularx}\\[0.2cm]
  \caption{ \label{table:method_v2}
    Overview of Green's function representation approaches in both Matsubara frequency space and imaginary time combined with the scaling of solvers for the Dyson equation. Where no convergence is listed, the scaling either involve additional parameters or is unknown.}
\end{table*}

In this paper, we present a Legendre spectral method for solving the Dyson equation with super exponential convergence and a convolution that scales quadratically $\sim \mathcal{O}(N_L^2)$, one order better than previous formulations\cite{PhysRevB.98.075127}.
The super exponential convergence allow us to achieve an energy accuracy of $10^{-9} E_h$ in a realistic quantum chemistry system with a few hundred expansion coefficients. We show this in a proof-of concept benchmark: computing the dissociation energy of He$_2$ using self-consistent second order perturbation theory, taking both the zero temperature and the complete basis limit.

This paper is organized as follows. In section~\ref{sect: general}, we introduce the Dyson equation. In section~\ref{sect: method}, we present our Legendre spectral method. In section~\ref{sect: application} and~\ref{sect: result}, we apply our method to a realistic quantum chemistry problem, the dissociation energy of the noble gas He$_2$. In section~\ref{sect: conclusion}, we present conclusions.

\section{Theory} \label{sect: general}

\begin{figure}
\includegraphics[scale=1] {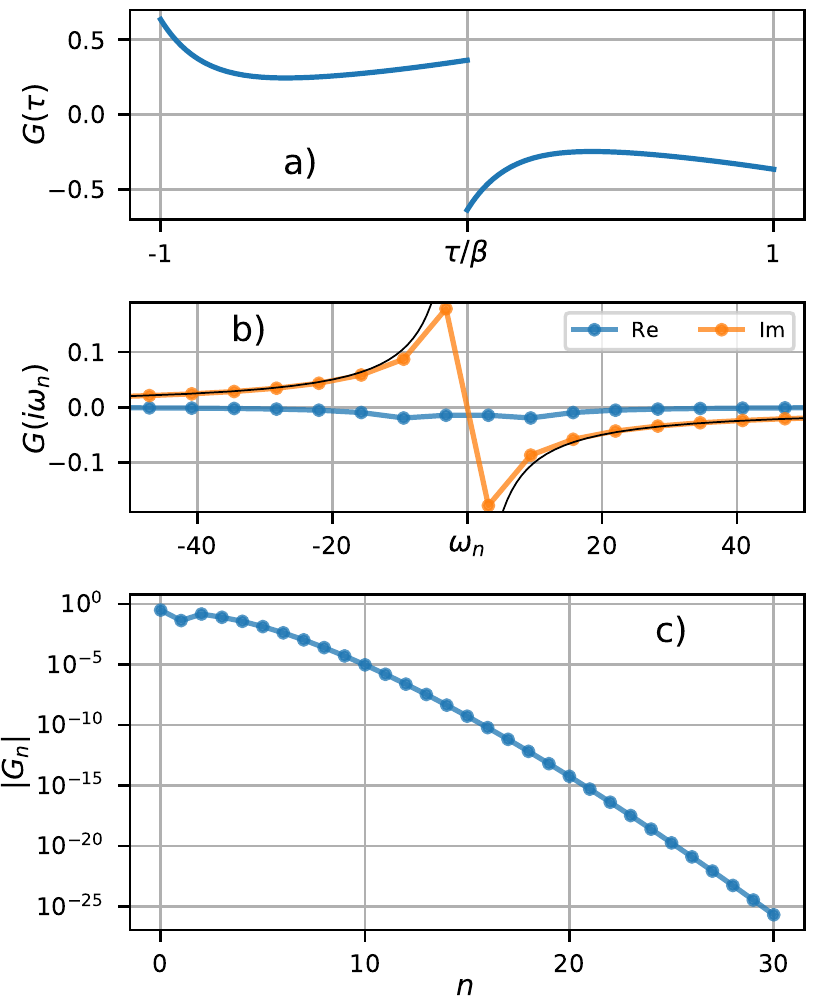} \\[-3mm]
\caption{\label{fig:g} Single particle Green's function in a) imaginary time $G(\tau)$, b) Matsubara frequency $G(i\omega_n)$ (with $(i\omega_n)^{-1}$ black line), and c) Legendre expansion coefficients $G_n$,
for site one in the fermionic two level system with the second quatization Hamiltonian $H = -\mu c^\dagger_1 c_1 + V ( c^\dagger_1 c_2 + c^\dagger_2 c_1) + \epsilon c^\dagger_2 c_2$ at inverse temperature $\beta = 1$, where $c^\dagger_i$ creates and $c_i$ annihilates a fermion at site $i$ and $\mu = -3$, $\epsilon = 3.3$, $V = 4$.
} \end{figure}

The imaginary time single particle Green's function $G$ is defined on the interval $\tau \in [-\beta, \beta]$, $G \equiv G(\tau)$, where $\beta$ is the inverse temperature $\beta = 1/T$. 
It obeys the periodicity condition $G(-\tau) = \xi G(\beta - \tau)$, with $\xi = +1$ ($-1$) for bosons (fermions), making it an (anti-)periodic function with a step discontinuity at $\tau = 0$, see Fig.~\ref{fig:g}a.
The imaginary time Dyson equation for $G(\tau)$ is \cite{Negele:1998aa, Fetter:2003aa, Atland:2006nx, Stefanucci:2013oq}
%
%
\begin{equation}
  [- \partial_\tau - h] G(\tau) - \Sigma \ast G = 0
  \, ,
  \label{eq:dyson}
\end{equation}
where $h$ is the single particle energy, and $\Sigma$ the self-energy, which accounts for all many-body interactions. We note in passing that $\Sigma(\tau)$ has the same periodicity as $G(\tau)$.
The boundary condition for Eq.~(\ref{eq:dyson}) is $G(0) - \xi G(\beta) = -1$,
%
%
and the Fredholm type \cite{Lax:2002aa} imaginary time convolution is defined as
$\Sigma \ast G \equiv \int_0^\beta d\bar{\tau} \, \Sigma(\tau - \bar{\tau}) G(\bar{\tau})$.
%
%

\begin{figure}
\includegraphics[scale=1] {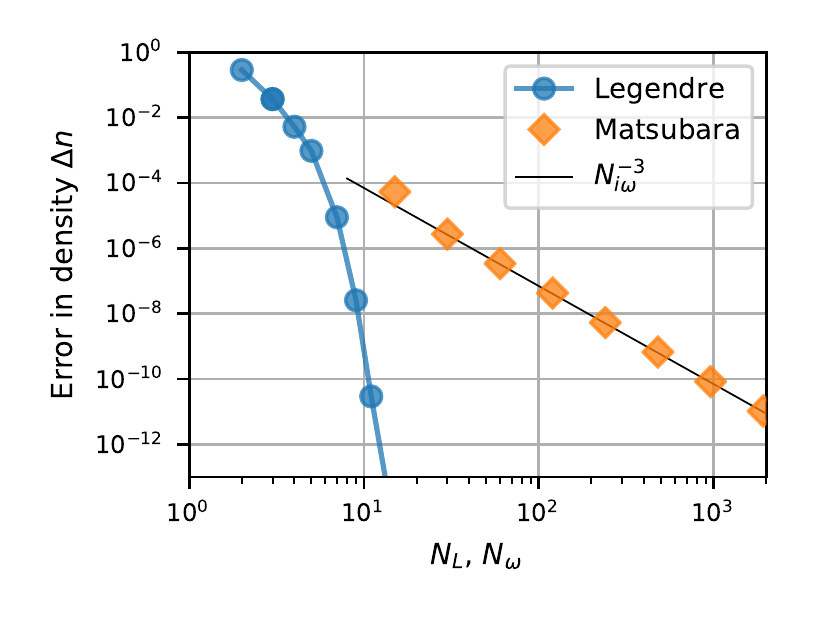} \\[-8mm]
\caption{\label{fig:density} Error in density $\Delta n$ as a function of Legendre expansion order $N_{L}$ and number of Matsubara frequencies $N_{\omega}$, for the same system as in Fig.~\ref{fig:g}.
} \end{figure}

Analytically the Dyson equation (Eq.~\ref{eq:dyson}) can be solved using the Fourier series expansion
\begin{equation*}
  G( \tau ) = \frac{1}{\beta} \!\! \sum_{n=-\infty}^{\infty} \!\!\!\! e^{-i\omega_n \tau} G(i \omega_n)
  \, , \ 
  G(i\omega_n) = \!\! \int_0^\beta \!\!\!\! d\tau \, e^{i\omega_n \tau} G(\tau),
\end{equation*}
%
where
the Matsubara frequencies $i\omega_n$ are given by $i\omega_n \equiv i \frac{\pi}{\beta} (2n + \eta)$ with $\eta = (1 - \xi) / 2$, and $n$ integers \cite{Negele:1998aa, Fetter:2003aa, Atland:2006nx, Stefanucci:2013oq}.
In Matsubara frequency space the Dyson equation (Eq.~\ref{eq:dyson}) is diagonal \cite{Matsubara:1955aa}
\begin{equation}
  [i\omega_n - h - \Sigma(i\omega_n)] G(i\omega_n) = 1
  \, .
\end{equation}
Numerically, however, the discontinuity at $\tau = 0$ results in a slow asymptotic decay $G(i\omega_n) \sim (i\omega_n)^{-1}$ as $i\omega_n \rightarrow \pm i\infty$, see Fig.~\ref{fig:g}b. 
This prevents a naive finite frequency $|n|<N_\omega$ approximation $G(\tau) \approx \frac{1}{\beta} \sum_{|n| < N_\omega} e^{-i\omega_n \tau} G(i\omega_n)$ from converging in $N_\omega$ (the maximal error in $G(\tau)$ scales as $\sim \mathcal{O}(N_\omega^0) = \mathcal{O}(1)$). The standard solution to this problem is to use a finite number $p$ of high-frequency "tail" coefficients $\bar{G}_k$ to approximate $G(i\omega_n) \approx \sum^p_{k=1} \bar{G}_k/(i\omega_n)^k$ for $|n| > N_\omega$, where the known asymptotic decay implies $\bar{G}_1 = 1$. This type of tail correction procedure gives polynomial convergence in $G(\tau)$ with the power determined by the order $p$ of the tail expansion $\sim \mathcal{O}(N_\omega^{-p})$, see e.g.\ Ref.~\onlinecite{Nils:2002aa, Comanac:2007aa, Dario:2016aa}.
%
%
%
%
%
%
In Fig.~\ref{fig:density} this is shown for the case of $p=3$ using the TRIQS library \cite{Parcollet2015398}. 

%
%
%

Since $G(\tau)$ is continuous on $\tau \in [0, \beta]$ it can be much more efficiently represented by a finite orthogonal polynomial expansion
\begin{equation}
  G(\tau) \approx \sum_{n=0}^{N_L} G_n L_n[x(\tau)]
  \, ,
  \label{eq:leg_exp}
\end{equation}
where $L_n[x]$ are Legendre polynomials defined on $x \in [-1,1]$ and $x(\tau) = \frac{2\tau}{\beta} - 1$.
%
%
The Legendre coefficients $G_n$ have a faster than exponential asymptotic decay \cite{Boehnke:2011fk}, see Fig.~\ref{fig:g}c. This also causes the finite $N_L$ expansion at the right hand side of Eq.~(\ref{eq:leg_exp}) to converge faster than exponential $\lesssim \mathcal{O}(e^{-N_L})$ to the analytical $G(\tau)$.

\section{Legendre spectral method}\label{sect: method}



Here we develop a Legendre spectral method for solving the Dyson equation (Eq.~\ref{eq:dyson}), 
reformulating the integro-differential equation in the space of Legendre coefficients $G_n$ (Eq.~\ref{eq:leg_exp}).
In the space of a finite Legendre expansion of order $N_L$, Eq.~(\ref{eq:dyson}) is cast to a linear equation system
\begin{equation}
  \sum_{n=0}^{N_L} ( - D_{kn} - h \mathbf{1}_{kn} - [\Sigma *]_{kn} ) \, G_n = \mathbf{0}_k
  \, ,
  \label{eq:linear_system}
\end{equation}
%
where terms with one and two indices are vectors and matrices in Legendre coefficient space.
The last row of the left hand side matrix is modified to enforce the boundary condition of Eq.~(\ref{eq:dyson}).
The resulting method has faster than exponential convergence and quadratic scaling $\sim \mathcal{O}( N_L^2 )$, one order better than previous approaches.\cite{PhysRevB.98.075127}

\begin{figure}
\includegraphics[scale=1.0] {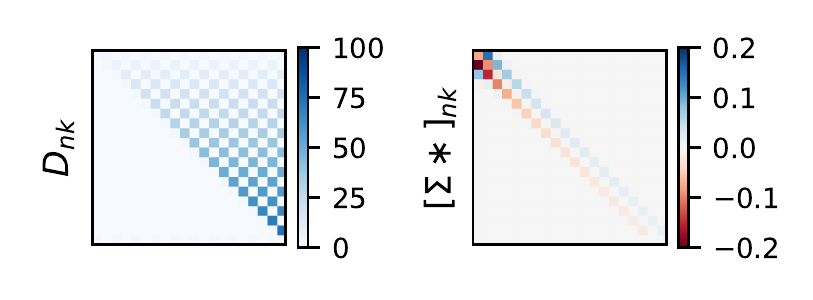} \\[-3mm]
\caption{\label{fig:spectral_matrices} (Color online) Matrix structure of
  the spectral derivative operator $D_{kn}$ and the convolution operator
  $[\Sigma \ast]_{kn}$ for $\epsilon = 1$ and $\Sigma(\tau) = e^{-\epsilon \tau} (\xi e^{-\epsilon \beta} - 1)^{-1}$ at
  $\beta = 1$ and $\xi = -1$ (fermions).
} \end{figure}

The differential operator $\partial_\tau$ in Eq.\ (\ref{eq:dyson}) acting on the Legendre polynomials takes the form \cite{Jie-Shen:2011uq}
\begin{multline}
  \partial_\tau L_n[x(\tau)]
  = \frac{2}{\beta} \partial_x L_n(x) \\
  = \frac{2}{\beta} \sum_{k=0, \, k+n \textrm{ odd}}^{n-1} (2k + 1) L_k(x) =
  \sum_k D_{kn} L_k(x)
  \, . 
  \label{eq:differential_op}
\end{multline}
%
%
Hence the derivative matrix $D_{kn}$ in Eq.~\ref{eq:linear_system} is given by
\begin{equation}
  \frac{\beta}{2} D_{kn}
  \equiv
  \left\{
  \begin{array}{lr}
    2k + 1, & 0 \le k \le n, k+n \textrm{ odd} \\
    0, & \textrm{elsewhere}
  \end{array}
  \right. \, ,
\end{equation}
and is upper triangular, see Fig.\ \ref{fig:spectral_matrices}.
%
%
%
Using $L_n(\pm 1) = (\pm 1)^n$ the Dyson equation boundary condition can be written as
\begin{equation}
  -1
  = G(0) - \xi G(\beta)
  =
  \sum_n ((-1)^n - \xi) \, G_n
  \, .
  \label{eq:bc}
\end{equation}
%
%

\subsection{Spectral convolution}


The imaginary time convolution $[\Sigma \ast G]$ in the Dyson equation (Eq.~\ref{eq:dyson}) can be separated into the two terms of Volterra type
\begin{align}
  & [\Sigma \ast G](\tau) =
  \int_0^\beta d\tau' \Sigma(\tau - \tau') G(\tau')
  \label{eq:volterra} \\ & =
  \int_0^\tau \!\! d\tau' \Sigma(\tau - \tau') G(\tau')
  + \int_\tau^\beta \!\! d\tau' \xi \Sigma(\beta + \tau - \tau')G(\tau')
  \, ,  \nonumber
\end{align}
using the periodicity property $\Sigma(-\tau) = \xi \Sigma(\beta - \tau)$. In Eq.~\ref{eq:volterra} $\Sigma(\tau)$ is only evaluated for $\tau \in [0, \beta]$, avoiding the discontinuity at $\tau = 0$.


%
%
%
In Legendre coefficient space the convolution operator $[\Sigma \, \ast]$ can be written as a sum of two matrices $B^\lessgtr_{kn}$ representing the two Volterra terms Eq.~(\ref{eq:volterra})
\begin{equation}
  [\Sigma \, \ast]_{kn} \equiv B^{<}_{kn} + \xi B^{>}_{kn}
  \, .
\end{equation}
Stable recursion relations for $B^\lessgtr_{nk}$ have been derived by Hale and Townsend \cite{Hale:2014ab} using the Fourier connection between Legendre polynomials and spherical Bessel functions.
Since the derivation is detailed in Ref.~\onlinecite{Hale:2014ab} we only state the result specialized to the imaginary time convolution in Eq.~(\ref{eq:volterra}) here, and provide a derivation in Appendix \ref{sect: Appendix} .

The coefficients are related by the recursion relation
\begin{equation}
  B^\lessgtr_{k, n+1} =
  - \frac{2n + 1}{2k + 3} B^\lessgtr_{k+1, n}
  + \frac{2n + 1}{2k - 1} B^\lessgtr_{k-1, n}
  + B^\lessgtr_{k, n-1}
  \label{eq:convol_recur}
\end{equation}
which for each column require two previous columns to be known. The recursion is only stable
for the lower triangular coefficients in $B^\lessgtr_{kn}$. The upper triangular coefficients
are computed using the transpose relation
\begin{equation}
  B^\lessgtr_{k,n} = (-1)^{n+k} \frac{2k + 1}{2n + 1} B^\lessgtr_{n,k}
  \, .
  \label{eq:convol_trans}
\end{equation}
The two first columns are given by the starting relations
\begin{multline}
  B^\lessgtr_{k, 0} =
    \left\{
  \begin{array}{lr}
    \Sigma_0 \pm \frac{\Sigma_1}{3} \, , & k = 0 \\[2mm]
    \pm (\frac{\Sigma_{k-1}}{2k - 1} - \frac{\Sigma_{k+1}}{2k + 3}) \, , & \quad k \geq 1
  \end{array}
  \right.
  \\
  B^\lessgtr_{k, 1} =
  \mp B^\lessgtr_{k, 0} + \frac{B^\lessgtr_{k-1, 0}}{2k - 1} - \frac{B^\lessgtr_{k+1, 0}}{2k + 3}
  \, , \quad k \ge 1
  \label{eq:starting}
\end{multline}
with the special case for $k = 0$, $B^\lessgtr_{0, 1} = \mp B^\lessgtr_{1, 0} / 3$,
using the Legendre coefficients $\Sigma_n$ of the self-energy $\Sigma$, c.f.\ Eq.~(\ref{eq:leg_exp}).\\

\subsection{Convergence and scaling}

Since each coefficient in $B^\lessgtr_{kn}$ can be computed in $\mathcal{O}(1)$ operations,
the scaling of the convolution matrix construction is $\sim \mathcal{O}(N_L^2)$.
The self-energy $\Sigma(\tau)$ is a smooth function with asymptotic exponentially decaying Legendre coefficients which causes the entries of the dominantly diagonal spectral convolution operator $[\Sigma \, \ast]_{kn}$ to decay exponentially both along and away from the diagonal, see Fig.\ \ref{fig:spectral_matrices}.



The numerical solution of $G(\tau)$ from the Dyson equation constructed in terms of the linear system in Eq.~(\ref{eq:linear_system}) converges faster than exponentially to the analytical solution, with increased number of Legendre coefficients $N_L$, see Fig.~\ref{fig:density}. This is in stark contrast to the polynomial convergence of the standard Matsubara tail approach \cite{Nils:2002aa, Comanac:2007aa, Dario:2016aa}, also shown in Fig.~\ref{fig:density}.

\subsection{Imaginary time transform}

To retain the high accuracy of the Legendre spectral Dyson solver the method has to be complemented with stable transforms between Legendre coefficients and imaginary time
\begin{equation}
  G_n = \sum_{i=0}^{N_L} S_{ni} G(\tau_i)
  \, , \quad
  G(\tau_i) = \sum_{n=0}^{N_L} L_{in} G_n
  \, .
\end{equation}
To construct the well-conditioned transform matrices $S_{ni}$ and $L_{in}$ we employ Legendre quadrature
and the Legendre-Gauss-Lobatto points $x_i \in \{ x : (1 - x^2) L_{N_L}(x)=0 \}$, $x_0=-1$, $x_N = 1$, re-scaled to the imaginary time interval $[0, \beta]$, $\tau_i = \beta \frac{x_i + 1}{2}$. Using $x_i$ the matrices $S_{ni}$ and $L_{in}$ can be directly constructed (avoiding matrix inversion)
\begin{equation}
  L_{in} = L_n( x (\tau_i) )
  \, , \quad
  S_{ni} = \frac{\beta}{2W_n} \omega_i L_n( x(\tau_i) ),
\end{equation}
where $\int_{-1}^{1} dx \, L_n(x) L_m(x) = \delta_{nm} \frac{2}{2n + 1} \equiv \delta_{nm} W_n$ and $ \omega_i = \frac{2}{N(N+1)} \frac{1}{L_{N_L}(x_i)^2}$, see e.g.\ Refs.~\onlinecite{Jie-Shen:2011uq,Jie-Shen:2006vn} .
\section{Application (GF2)}\label{sect: application}



As a proof of concept application of the Legendre spectral Dyson solver developed in this paper we employ the solver in a quantum chemistry setting using a Gaussian basis set.
We will employ self-consistent second order perturbation theory, also known as GF2 \cite{PhysRevB.63.075112, Dahlen:2005aa, Phillips:2014aa, Phillips:2015aa, Kananenka:2016aa, Kananenka:2016ab, Rusakov:2016aa, Welden:2016aa, Iskakov:2019aa}, which has seen a revival in recent years, both in \textit{ab initio} condensed matter applications \cite{doi:10.1021/ct5001268, Rusakov:2016aa} and in quantum chemistry \cite{Phillips:2014aa, Welden:2016aa, PhysRevB.97.115164, Neuhauser:2017aa} in combination with embedding methods \cite{Zgid:2017aa}. Our implementation is built on the Coulomb integrals of the pyscf library \cite{Sun:2018aa}.
%

In the resulting non-orthogonal basis set the Dyson equation takes the form
\begin{equation}
  \sum_j [S_{ij}( \partial_\tau - \mu ) + F_{ij} + \Sigma_{ij} \ast ] \,  G_{jk}(\tau) = \mathbf{0}
  \label{eq:dysonS}
\end{equation}
in which $i, j, k$ are orbital indices, $S_{ij}$ is the overlap matrix, and $F_{ij}$ is the so called Fock matrix, $F_{ij} \equiv h_{ij} + \Sigma^{\text{(HF)}}_{ij}$. The boundary condition for this equation is $\sum_j (G_{ij}(0) - \xi G_{ij}(\beta)) \cdot S_{jk} = -\mathbf{1}_{ik}$. Here, the single particle term $h_{ij}$ accounts for electronic kinetic and nuclear-electronic matrix elements and 
the Hartree-Fock self energy $\Sigma^{\text{(HF)}}_{ij}$ is given by
%
%
\begin{equation}
    \Sigma^\text{({HF})}_{ij} = \sum_{kl}P_{kl}(v_{ijkl} - v_{ilkj}/2)
    \, , \label{eq:HF}
\end{equation}
were $P_{ij}$ the density matrix $P_{ij} = -2G_{ij}(\beta)$, and $v_{ijkl}$ the electron-electron Coulomb repulsion integral.
%

%
%
In GF2 the imaginary-time-dependent part of the self energy $\Sigma(\tau)$ is approximated with the second order self energy diagram using the full electron Greens function $G$, $\Sigma \approx \Sigma^{\text{(GF2)}} [ G ]$ where
\begin{equation}
    \begin{aligned}
        \Sigma_{ij}^\text{(GF2)}(\tau) = \sum_{klmnpq} G_{kl}(\tau) &G_{mn}(\tau) G_{pq}(\beta - \tau) \\ \times &v_{impk}(2v_{jnlq} - v_{jlnq})
        \, .
    \end{aligned}
    \label{eq:GF2}
\end{equation}
The evaluation of $\Sigma^{\text{(GF2)}}(\tau)$ for fixed $\tau$ scales as $\sim \mathcal{O}(N^5)$ \cite{Neuhauser:2017aa}, where $N$ is the number of atomic orbitals.
%

%
%

\begin{figure}
\includegraphics[scale=1] {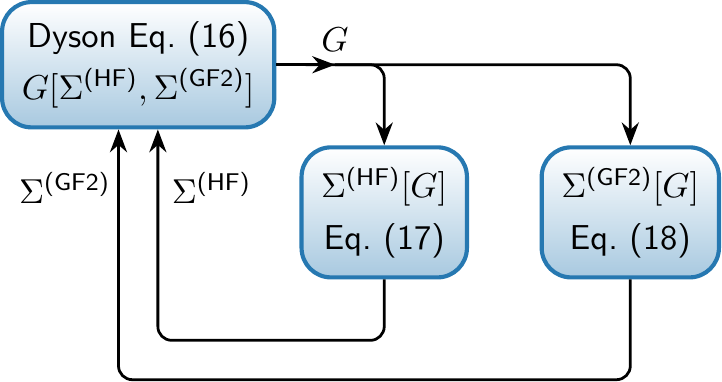} \\[-0mm]
\caption{\label{fig:GF2} Schematic GF2 self consistency loop.
} \end{figure}

Solving for the GF2 Greens function $G$ amounts to solving Eqs.~(\ref{eq:dysonS}), (\ref{eq:HF}), and (\ref{eq:GF2}) which is a highly non-linear problem.
%
%
To find the solution we perform self-consistent iterations, see Fig.~\ref{fig:GF2} for a schematic picture. The inner loop solves the Dyson equation [Eq.~(\ref{eq:dysonS})] and updates the Hartree-Fock self energy $\Sigma^{\text{(HF)}}$ [Eq.~(\ref{eq:HF})] until convergence (in the Fock-matrix $F$). At convergence in $F$, one step of the outer loop is performed by re-evaluating the GF2 self energy $\Sigma^{\text{(GF2)}}$ [Eq.~(\ref{eq:GF2})] and computing the relative change in total energy $E$. If the change is above a fixed threshold, the inner loop is started again. To compute the inter molecular energies, which is an energy difference, we need a threshold of $10^{-10}$.
%

The total energy $E$ of the system is given by
\begin{equation}
    E = \frac{1}{2} \text{Tr}[(h + F)P] + \text{Tr}[\Sigma * G] + E_{nn}
    \, ,
    \label{eq:E_GF2}
\end{equation}
%
%
where $E_{nn}$ is the nuclei-nuclei Coulomb energy.
The imaginary time trace $\text{Tr}[\cdot]$ is defined as $\text{Tr}[A] \equiv -\sum_{i}A_{ii}(\beta)$ \cite{Dario:2016aa}\textsuperscript{,}\footnote{This follows from the generalized imaginary time trace
$\textrm{Tr}[ G ] \equiv \frac{-\xi}{\beta} \sum_{ab} \iint_0^\beta \, d\tau d\tau' \, \delta_{a,b} \delta(\tau - \tau' + 0^-) G_{ab}(\tau, \tau')$
of a response function
$G_{ab}(\tau, \tau') \equiv - < \! \mathcal{T} c_a(\tau) c^\dagger_b(\tau') \! >$.
} and the $\Sigma * G$ convolution is computed with the spectral Legendre convolution as in Eq.~(\ref{eq:linear_system}).
%

\section{Results}\label{sect: result}

The faster than exponential convergence of the Legendre spectral Dyson solver Eq.~(\ref{eq:linear_system}) is particularly suited for high precision calculations. A prime example is the computation of the binding energy $D_e$ in noble-gas dimers, where the weak bonding requires high precision calculations of total energies.
The binding energy $D_e$ is obtained from the the minimum of the interaction energy $E_{\text{int}}(r)$ as a function of atomic separation $r$
\begin{equation}
  D_e \equiv - E_{\text{int}}(r_e) \equiv - \min_r E_{\text{int}}(r)
  \, ,
\end{equation}
where $r_e$ is the equilibrium atomic distance.
The interaction energy $E_{\text{int}}$ is in turn given by
\begin{equation}
  E_{\text{int}}(r) \equiv E_{A_2}(r) - 2 E_{A}(r)
  \, ,
\end{equation}
where $E_{A_2}$ is the total energy of the dimer and $E_{A}$ is the total energy of the single atom (the monomer) evaluated using the standard counterpoise correction \cite{Boys:1970}.
In the noble gases the total energies $E_A$ and $E_{A_2}$ are of the order of Hartrees ($\sim E_h \equiv 1\,$Ha) while the binding energy $D_e$ is of the order of tens of micro Hartrees ($\sim 10\,\mu E_h$), hence requiring high precision calculation of the total energies.



We use He$_2$ as a prototype system since there exist published reference results for the binding energy $D_e$ and equilibrium distance $r_e$ calculated with Hartree-Fock (HF), second-order Moller–Plesset perturbation theory (MP2), coupled cluster singles doubles (CCSD) theory and coupled cluster singles doubles and non-iterative perturbative triples (CCSD(T)) theory \cite{Van-Mourik:1999aa}.
%
%
%
The MP2 method is closely related to GF2 and uses the second order self energy [Eq.~(\ref{eq:GF2})] evaluated at the HF Green's function $G^{\text{(HF)}}$, $\Sigma^{\text{(MP2)}} \equiv \Sigma^{\text{(GF2})}[ G^{\text{(HF)}} ]$. Note however that the prefactors in the total energy differ \cite{Holleboom:1990aa, Dahlen:2005aa}.

\begin{figure}[h!]
\includegraphics[scale=1] {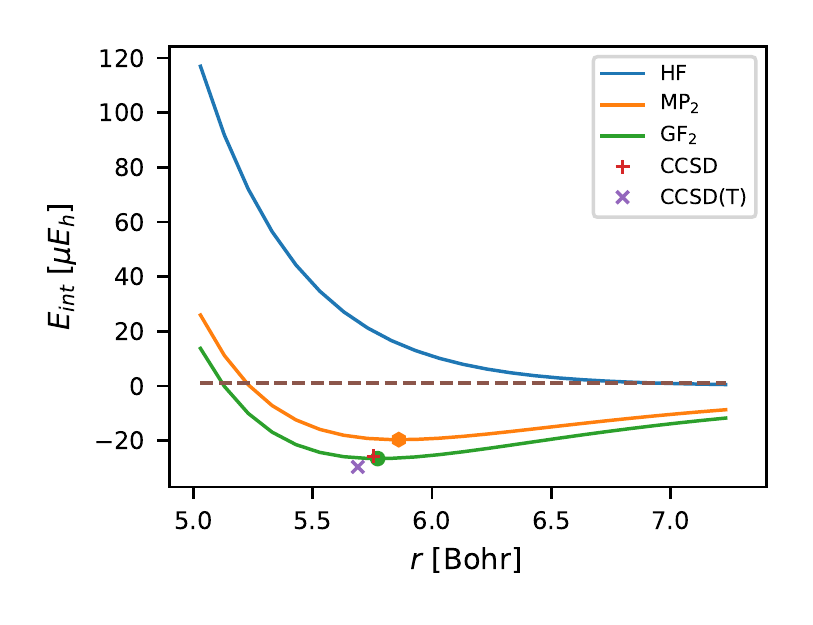} \\[-3mm]
\caption{\label{fig:aug-ccpvqz} Interaction energy $E_{\text{int}}$ as a function of atomic distance $r$ of He$_2$ with basis aug-ccpvqz using HF, MP2 and GF2. The HF and MP2 results are computed with pyscf \cite{Sun:2018aa}, the GF2 results are computed using $\beta = 50\,E_h^{-1}$, $N_\tau = 192$. The CCSD and CCSD(T) results are from Ref.~\onlinecite{Van-Mourik:1999aa}.
} 
\end{figure}

Fig.~\ref{fig:aug-ccpvqz} shows $E_{\text{int}}(r)$ (and $-D_e$) of He$_2$ computed with HF, MP2, and GF2 in the aug-cc-pvqz basis together with CCSD and CCSD(T) reference results on $D_e$ \cite{Van-Mourik:1999aa}.
The GF2 results are obtained by fitting a 4th order polynomial to 21 $r$-points of $E_{\text{int}}(r)$ computed in a $0.1\,$ Bohr range centered around the minimum at $r_e$.
The GF2 results are obtained using the Legendre spectral Dyson solver while HF and MP2 are computed using pyscf \cite{Sun:2018aa}.
%
%
%
As seen in Fig.~\ref{fig:aug-ccpvqz} He$_2$ does not bind within the Hartree-Fock approximation which gives a strictly positive interaction energy. Compared to MP2 our GF2 results are a considerable improvement, using the coupled cluster CCSD and CCSD(T) as reference.

\subsection{Complete basis set limit}

In order to extrapolate the results to the complete basis set (CBS) limit \cite{Feller1992:aa,Helgaker1997:aa} we repeat the calculations using the augmented correlation consistent (aug-cc-pv$n$z) basis set series with $n=$ d, t, q, 5 (i.e.\ $n=2,3,4,5$) \cite{Kendall:1992aa, Woon:1993aa, Woon:1994aa}.
This series has been shown to enable accurate extrapolation of a number of properties due to its systematic convergence in $n$ \cite{Feller1992:aa, Woon:1993ab, Woon:1993ac, Woon:1993ad, Peterson:1993aa, Xantheas:1993aa, Woon:1994ab, Woon:1994ac, Woon:1996aa, Mourik:1997aa, Peterson:1997aa,Peterson:1997ab}.
%

\begin{table}
  \begin{tabular}{ l | c c c c}
    $D_e$ [$\mu E_h$]
    & $\,$ MP2 $\,$ & $\,$ CCSD $\,$ & CCSD(T) & $\,$ GF2 $\,$ \\
    \hline
    aug-ccpvdz & 12.69 & 16.78 & 18.57 & 18.17 \\
    aug-ccpvtz & 17.97 & 23.77 & 27.10 & 24.63 \\
    aug-ccpvqz & 19.66 & 25.79 & 29.64 & 26.59 \\
    aug-ccpv5z & 20.71 & 27.09 & 31.25 & 27.79 \\
    CBS & 22.98 & 30.06 & 34.70 & 29.67 \\
  \end{tabular}\\[0.2cm]
  \begin{tabular}{ l | c c c c}
    $r_e$ [Bohr]
    & $\,$ MP2 $\,$ & $\,$ CCSD $\,$ & CCSD(T) & $\,$ GF2 $\,$ \\
    \hline
    aug-ccpvdz & 6.1680 & 6.0580 & 6.0086 & 6.0547 \\
    aug-ccpvtz & 5.9175 & 5.8060 & 5.7452 & 5.8244 \\
    aug-ccpvqz & 5.8606 & 5.7546 & 5.6891 & 5.7722 \\
    aug-ccpv5z & 5.8244 & 5.7210 & 5.6537 & 5.7388 \\
    CBS & 5.769 & 5.672 & 5.607 & 5.680 \\
  \end{tabular}  
\caption{\label{table:De_re}
  Dissociation energies $D_e$ (top) and Equilibrium distances $r_e$ computed by MP2, CCSD, CCSD(T), and GF2 with the basis sets aug-cc-pv$n$z, with $n=$d, t, q, 5. The MP2, CCSD and CCSD(T) results are from Ref.\ \onlinecite{Van-Mourik:1999aa}.
}
\end{table}

In Tab.~\ref{table:De_re}, we summarize the binding energy $D_e$ and equilibrium distance $r_e$ of He$_2$ computed by MP2, CCSD, CCSD(T) and GF2 using the aug-cc-pv$\{$d,t,q,5$\}$z basis sets.
The aug-cc-pv$\{$d,t,q,5$\}$z GF2 energies are computed at $\beta = 50\,E_h^{-1}$ using $N_\tau = 128$, 160, 192, and 250 $\tau$-points, respectively.
%
%
The convergence in $N_\tau$ is imposed so that the absolute values of the elements in highest Legendre coefficient matrix are smaller than $10^{-10}$.
%
%
The zero temperature convergence (at $\beta = 50\,E_h^{-1}$) is ensured by requiring that the finite temperature MP2 total energy differ with less than 0.1 nano Hartree compared to the zero temperature MP2 total energy from pyscf.


We note that the number of $\tau$-points $N_\tau$ used for the aug-cc-pv$\{$d,t,q,5$\}$z basis sets are of the same order as the number of atomic orbitals $N$. Hence, the scaling of GF2, $\sim \mathcal{O}(N_{\tau} \cdot N^5)$, is comparable to the scaling of CCSD, $\sim \mathcal{O}(N^6)$.
As seen in Tab.~\ref{table:De_re}, the accuracy of the GF2 result for $D_e$ is comparable to CCSD when compared to CCSD(T), while the CCSD result for $r_e$ is closer to CCSD(T) result than GF2. This makes GF2 a considerable improvement over MP2.

\begin{figure}
\includegraphics[scale=1] {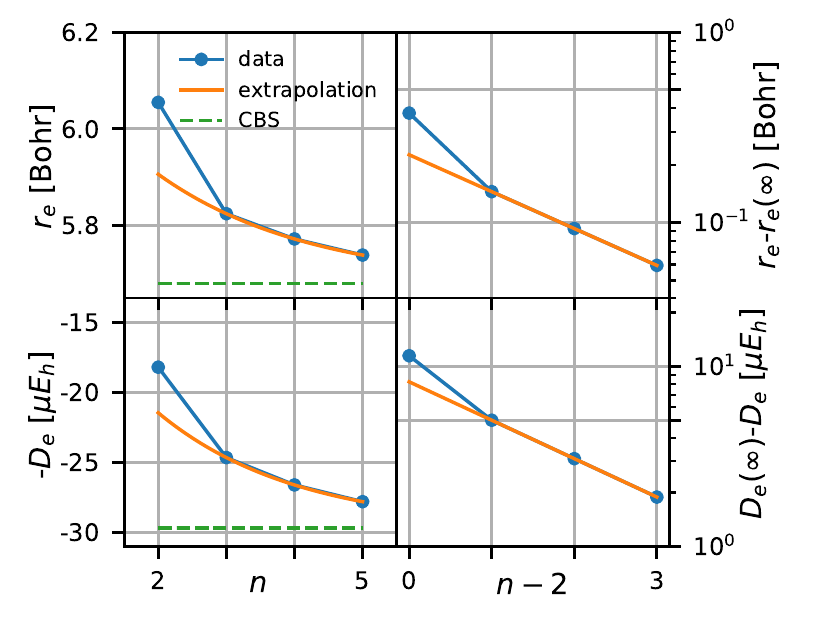} \\[-3mm]
\caption{\label{fig:extra} Basis extrapolation of equilibrium distance ($r_e$) and dissociation energy ($D_e$) He$_2$ with basis aug-ccpvnz with n = 2, 3, 4, 5. Left panels: raw data and fitting. Right panels: check of fitting results. 
} \end{figure}

%
With the systematic convergence of $D_e$ and $r_e$ as a function of basis set size $n$ it is possible to extrapolate to the complete basis limit $n \rightarrow \infty$ \cite{Van-Mourik:1999aa}.
%
%
We extrapolate $D_e$ and $r_e$ using our GF2 aug-ccpv\{t,q,5\}z results by fitting the exponential model: $A \cdot e^{-B(n - 2)} + C$, proposed in Ref.~\onlinecite{Van-Mourik:1999aa}, where $A$, $B$, and $C$ are parameters.
%
%
The applicability of the model is checked by a logarithmic plot, see Fig.~\ref{fig:extra}.
The resulting CBS limit of our GF2 results are $D_e \approx 29.67\,\mu E_h$ and $r_e \approx 5.680\,a_0$, see also Tab.~\ref{table:De_re}.

\section{Conclusion and outlook}\label{sect: conclusion}

%
We introduce a Legendre-spectral algorithm for solving the Dyson equation in Legendre coefficient space.
%
%
By staying in Legendre-coefficient space the algorithm converges super exponentially with respect to the number of Legendre coefficients $N_L$ used to represent the imaginary time Green's function \cite{Boehnke:2011fk}. This is in stark contrast to the Matsubara frequency space based approach with only polynomial convergence \cite{Nils:2002aa, Comanac:2007aa, Dario:2016aa}.
%
%
%
The exponential convergence is shared with a recently presented Chebyshev polynomial based algorithm \cite{PhysRevB.98.075127}, where the convolution scales as $\sim$ $\mathcal{O}(N_L^3)$.
Currently there is no known algorithm for Chebyshev series that can compute the convolution term with the same efficiency as in the Legendre series \cite{Hale:2014ab}. 
Our work goes beyond this, employing a Legendre convolution with $\mathcal{O}(N_L^2)$ scaling, enabling the application to larger \textit{ab initio} systems.

To benchmark the algorithm we apply it to the quantum chemistry computation of the dissociation energy of the noble gas He$_2$ using self-consistent second order perturbation theory (GF2). The exponential convergence of our algorithm allows us to reach the required $10^{-9} E_h$ zero temperature total-energy precision using only $100 - 200$ Legendre coefficients in the Dunning basis series aug-ccpv$n$z \cite{Kendall:1992aa, Woon:1993aa, Woon:1994aa}.

%
%
%
%
The algorithm is also relevant for condensed matter \textit{ab initio} applications in periodic systems that require high precision, such as GF2  \cite{PhysRevB.63.075112,Dahlen:2005aa, Phillips:2014aa, Phillips:2015aa, Kananenka:2016aa, Kananenka:2016ab, Rusakov:2016aa, Welden:2016aa, Iskakov:2019aa} and Hedin's $GW$ \cite{Hedin:1965aa, Aryasetiawan:1998aa, Stan:2009aa, Kutepov:2009aa, vanSetten:2015aa, Maggio:2017aa, Grumet:2018aa, Kutepov:2016aa, Kutepov:2017aa}.
This is a promising venue for future research.

\begin{acknowledgments}

HURS would like to acknowledge helpful discussions and support from
i) Alex Barnett and Manas Rachh on the fundamentals of spectral methods
ii) Keaton Burns for pointing out Ref.~\onlinecite{Hale:2014ab}
iii) Sergei Iskakov for providing independent reference results for testing the pyscf-GF2 implementation and useful discussions
vi) Lewin Boehnke, Andreas Herrmann, Philipp Werner, and Hiroshi Shinaoka, who took part in early discussion on Green's function representations which guided the development of the method
and v) Antoine Georges, Olivier Parcollet, Manuel Zingl, Alexandru Georgescu, Igor Krivenko, and Nils Wentzell, who contributed both through discussions and contributions to the TRIQS project.
EG and XD were supported by the Simons Collaboration on the many-electron problem.
DZ acknowledges support from NSFCHE-1453894.
The Flatiron Institute is a division of the Simons Foundation.
\end{acknowledgments}

\appendix
\section{Convolution matrix} \label{sect: Appendix}
In this appendix, we derive Eqs.~\ref{eq:convol_recur},~\ref{eq:convol_trans},~\ref{eq:starting} in the main text. The derivation follows Ref.~\onlinecite{Hale:2014ab} but with more details for both integrals in Eq.~\ref{eq:volterra}.

\subsection{Convolution and Fourier transform}
The convolution of two continuous integrable functions is defined as~\cite{Hale:2014ab}
\begin{equation}
  h(x) = (f \ast g)(x) \equiv \int_{-\infty}^\infty dt \, f(t) g(x - t)
  \, .
  \label{eq:convol}
\end{equation}
With the assumption $f$ and $g$ are periodic functions, their Fourier transform can be written as 
\begin{equation}
  \mathcal{F}\{ f \} (\omega) = \int_{-\infty}^\infty dx \, e^{-i\omega x} f(x) 
  \, ,
\end{equation}

\begin{equation}
  \mathcal{F}^{-1}\{ f \} (x) = \frac{1}{2\pi} \int_{-\infty}^\infty dx \, e^{i\omega x} f(x)
  \, ,
\end{equation}
which satisfy the Fourier inversion theorem $\mathcal{F}^{-1} \{ \mathcal{F} \{ f \} \} = f$ and convolution theorem~\cite{Katznelson:1976aa}
\begin{equation}
  \mathcal{F} \{ f \ast g \} = \mathcal{F} \{ f \} \cdot \mathcal{F} \{ g \}
  \, .
  \label{eq:convolution}
\end{equation}

\subsection{Legendre polynomials}
The Legendre polynomials $P_n(x)$ can be defined recursively using the three term recurrence relation
\begin{multline}
  P_0(x) = 1, \quad P_1(x) = x, \\
  (n + 1) P_{n+1}(x) = (2n + 1) x P_n(x) - n P_{n-1}(x),
\end{multline}
They are orthogonal on $[-1, 1]$
\begin{equation}
  \int_{-1}^1 dx \, P_m(x) P_n(x) = \delta_{m,n} \frac{2}{2n + 1}
  \label{eq:LegOverlap}
\end{equation}
and the derivatives satisfy the recurrence relation
\begin{equation}
  (2n + 1) P_n(x) = \frac{d}{dx} \left[ P_{n+1}(x) - P_{n-1}(x) \right]
  \label{eq:LegDer}
\end{equation}
The Fourier transform and inverse Fourier transform of the Legendre polynomials can be expressed in terms of Bessel functions of the first kind
\begin{equation}
  \mathcal{F} \{ P_n \} = \int_{-1}^1 dx \, e^{-i \omega x}P_n(x) = 2(-i)^n j_n(\omega)
  \label{eq:Fourier}
\end{equation}
\begin{equation}
  \mathcal{F}^{-1} \{ P_n \} = \int_{-1}^1 dx \, e^{i \omega x}P_n(x) = 2i^n j_n(\omega)
\end{equation}
where $j_n(z)$ is the $n$th spherical Bessel function, and $P_n = 0$ outside $[-1, 1]$.\\
By combining Eq.~(\ref{eq:convolution}) and Eq.~ (\ref{eq:Fourier}), the convolution of Legendre polynomials can be expressed in terms of Bessel functions
\begin{equation}
  (P_m \ast P_n)(x) = \frac{2(-i)^{m + n}}{\pi} \int_{-\infty}^\infty d\omega \,
  e^{i\omega x} j_m(\omega) j_n(\omega)
  \label{eq:convol_fourier}
\end{equation}
This is the central observation of Ref.~\onlinecite{Hale:2014ab} that enables the derivation of recursion relations for the Legendre polynomial convolution.

The main property of Spherical Bessel functions used is the three term recurrence relation
\begin{multline}
  j_{-1}(z) = \frac{\cos z}{z}
  \, , \quad
  j_0(z) = \frac{\sin z}{z}
  \, , \\
  j_{n+1}(z) = \frac{2n + 1}{z} j_n(z) - j_{n-1}(z)
  \, , \quad
  n \ge 0
  \label{eq:bessel_recursion}
\end{multline}


The convolution equation Eq.~(\ref{eq:convol}) can be computed by replacing the two continuous function $f(x)$ and $g(x)$ on bounded interval with polynomial approximates $f_M(x)$ and $g_N(x)$ of sufficiently high degree.
With two Legendre series $f_M(x)$ and $g_N(x)$ supported on $x \in [-1, 1]$
\begin{equation}
  f_M(x) = \sum_{m=0}^{M} \alpha_m P_m(x)
  \, , \quad
  g_N(x) = \sum_{n=0}^N \beta_n P_n(x)
  \, ,
\end{equation}
Eq.~(\ref{eq:convol}) becomes
\begin{multline}
  h(x) = (f_M \ast g_N)(x) =
  \int_{\max(-1,x-1)}^{\min(1,x+1)} dt \,
  f_M(t) g_N(x - t)
  \\= \int_{-1}^{x+1} dt \,
  f_M(t) g_N(x - t) + \int_{x-1}^{1} dt \,
  f_M(t) g_N(x - t)
  \, ,
\end{multline}
which can be computed separately in two integration domain $x \in [-2, 0]$ and $x \in [0, 2]$ (see Fig. 4.1 in Ref.~\onlinecite{Hale:2014ab}).

\subsubsection{First interval $x \in [-2, 0]$}

For $x \in [-2, 0]$ we have $h(x) = h^<(x)$ where
\begin{equation}
  h^<(x) = \int_{-1}^{x+1} dt f_M(t) g_N(x - t) = \sum_{k=0}^{M + N + 1} \gamma_k^< P_k(x + 1) \, .
\end{equation}
Using the orthogonality of Legendre polynomials Eq.~(\ref{eq:LegOverlap}), we have 
\begin{equation}
\begin{aligned}
  \gamma_k^< &=
  \frac{2k + 1}{2}
  \int_{-2}^0 dx \, P_k(x + 1) \int_{-1}^{x+1} dt \, f_M(t) g_N(x - t)
  \\ &=
  \sum_{n=0}^N \beta_n
  \frac{2k + 1}{2}
  \sum_{m=0}^M \alpha_m 
  \\ &\times
  \int_{-2}^0 dx \, P_k(x + 1)
  \int_{-1}^{x+1} dt \, P_m(t) P_n(x - t)
  \label{eq:gamma_less}
\end{aligned}
\end{equation}
collecting terms we can write $\gamma_k^< = \sum_{n=0}^N B_{k.n}^< \beta_n$ where
\begin{equation}
\begin{aligned}
  B^<_{k, n} &=
  \frac{2k + 1}{2}
  \sum_{m=0}^M \alpha_m 
  \\ &\times
  \int_{-2}^0 dx \, P_k(x + 1)
  \int_{-1}^{x+1} dt \, P_m(t) P_n(x - t)
  \\ &=
  \frac{2k + 1}{2}
  \sum_{m=0}^M \alpha_m
  \int_{-2}^0 dx \, P_k(x + 1)
  (P_m \ast P_n) (x)
  \\ &=
  \frac{2k + 1}{2}
  \sum_{m=0}^M \alpha_m
  \int_{-1}^1 ds \, P_k(s)
  (P_m \ast P_n) (s - 1)
  \, .
\end{aligned}
\end{equation}
Using the Fourier expression for the Legendre convolution (Eq.~(\ref{eq:convol_fourier})), $B^<_{k, n}$ can be expressed in terms of spherical Bessel functions
\begin{multline}
  B^<_{k, n} =
  \frac{2k + 1}{\pi}
  \sum_{m=0}^M
  (-i)^{m+n}
  \alpha_m
  \\ \times
  \int_{-1}^1 ds \, P_k(s)
  \int_{-\infty}^\infty
  d\omega \,
  e^{i\omega(s - 1)}
  j_m(\omega) j_n(\omega)
  \, .
\end{multline}
Consider the $B^<_{k, n + 1}$ term, changing the order of integration and Fourier transforming the remaining Legendre polynomial gives
\begin{multline}
  B^<_{k, n+1} =
  \frac{2(2k + 1)}{\pi}
  \sum_{m=0}^M
  (-i)^{m+n+1} \, i^k
  \alpha_m
  \\ \times
  \int_{-\infty}^\infty
  d\omega \,
  j_k(\omega)
  j_m(\omega) j_{n+1}(\omega)
  e^{-i\omega}
  \, .
  \label{eq:B_kn+1}
\end{multline}
Applying the recursion relation of the spherical Bessel functions (Eq.~(\ref{eq:bessel_recursion})) on $n$ and $k$, we have
\begin{multline}
  (-i)^{m + n + 1} i^k j_k(\omega) j_m(\omega) j_{n+1}(\omega)
  \\ =
  (-i)^{m + n + 1} i^k j_k(\omega) j_m(\omega)
  \left(
  \frac{2n + 1}{\omega} j_n(\omega) - j_{n-1}(\omega)
  \right)
  \\ =
  \frac{2n + 1}{2k + 1} (-i)^{m + n + 1} i^k
  \left( j_{k + 1}(\omega) + j_{k - 1}(\omega) \right) j_m(\omega) j_n(\omega)
  \\
  + (-i)^{m + n - 1} i^k j_k(\omega) j_m(\omega) j_{n-1}(\omega)
\end{multline}
Back insertion in Eq.~(\ref{eq:B_kn+1}) and simplifying prefactors in $k$ gives
\begin{equation}
  B^<_{k, n+1} =
  - \frac{2n + 1}{2k + 3} B^<_{k+1, n}
  + \frac{2n + 1}{2k - 1} B^<_{k-1, n}
  + B^<_{k, n-1}
\end{equation}

\subsubsection{Second interval $x \in [0, 2]$}

For $x \in [0, 2]$ we have $h(x) = h^>(x)$ where

\begin{equation}
  h^>(x) =
  \int_{x-1}^1 dt \,
  f_M(t) g_N(x - t)
  = \sum_{k=0}^{M + N + 1} \gamma_k^> P_k(x - 1)
  \, .
\end{equation}
$\gamma_k^>$ can be computed in the same way as $\gamma_k^<$, see Eq.~(\ref{eq:gamma_less})
\begin{equation}
\begin{aligned}
  \gamma_k^> &=
  \frac{2k + 1}{2}
  \int_{0}^2 dx \, P_k(x - 1) \int_{x-1}^{1} dt \, f_M(t) g_N(x - t)
  \\ &=
  \sum_{n=0}^N \beta_n
  \frac{2k + 1}{2}
  \sum_{m=0}^M \alpha_m 
  \\ &\times
  \int_{0}^2 dx \, P_k(x - 1)
  \int_{x-1}^{1} dt \, P_m(t) P_n(x - t)
\end{aligned}
\end{equation}
collecting terms we can write $\gamma_k^> = \sum_{n=0}^N B_{k.n}^> \beta_n$ where
\begin{equation}
\begin{aligned}
  B^>_{k, n} &=
  \frac{2k + 1}{2}
  \sum_{m=0}^M \alpha_m 
  \\&\times 
  \int_{0}^2 dx \, P_k(x - 1)
  \int_{x-1}^{1} dt \, P_m(t) P_n(x - t)
  \\ &=
  \frac{2k + 1}{2}
  \sum_{m=0}^M \alpha_m
  \int_{0}^2 dx \, P_k(x - 1)
  (P_m \ast P_n) (x)
  \\ &=
  \frac{2k + 1}{2}
  \sum_{m=0}^M \alpha_m
  \int_{-1}^1 ds \, P_k(s)
  (P_m \ast P_n) (s + 1)
\end{aligned}
\end{equation}
using the Fourier expression for the Legendre convolution (Eq.~(\ref{eq:convol_fourier})) gives
\begin{multline}
  B^>_{k, n} =
  \frac{2k + 1}{\pi}
  \sum_{m=0}^M
  (-i)^{m+n}
  \alpha_m
  \\ \times
  \int_{-1}^1 ds \, P_k(s)
  \int_{-\infty}^\infty
  d\omega \,
  e^{i\omega(s + 1)}
  j_m(\omega) j_n(\omega)
\end{multline}
Since the exponent in the integral is unchanged when applying the recursion relations of the spherical Bessel functions we conclude that $B^>$ obeys the same recursion relation as $B^<$, albeit with a different starting point since the ``seeding'' integrals have a different sign in the exponent.

\subsubsection{Summary}

The convolution matrices for both the intervals can be expressed as the integral sums
\begin{multline}
  B^\lessgtr_{k, n} =
  \frac{2(2k + 1)}{\pi}
  \sum_{m=0}^M
  (-i)^{m+n} \, i^k
  \alpha_m
  \\ \times
  \int_{-\infty}^\infty
  d\omega \,
  j_k(\omega)
  j_m(\omega) j_n(\omega)
  e^{\mp i\omega}
\end{multline}
differing only in the sign in the exponent.
The coefficients are related by the recursion relation
\begin{equation}
  B^\lessgtr_{k, n+1} =
  - \frac{2n + 1}{2k + 3} B^\lessgtr_{k+1, n}
  + \frac{2n + 1}{2k - 1} B^\lessgtr_{k-1, n}
  + B^\lessgtr_{k, n-1}
\end{equation}
In practice this recursion relation is only stable below the diagonal with $k > n$. To get entries above diagonal, the transpose relation, that can be derived from the integral expression Eq.~\ref{eq:B_kn+1}, is used
\begin{equation}
  B^\lessgtr_{k,n} = (-1)^{n+k} \frac{2k + 1}{2n + 1} B^\lessgtr_{n,k}
\end{equation}

\subsection{Initial values $B^\lessgtr_{k,0}$ and $B^\lessgtr_{k,1}$}

To start the recursion, the initial values for $n= 0$ and $1$ are needed. To derive explicit expressions for these terms we repeatedly use the Volterra integral formula for Legendre polynomials from Ref.\ \onlinecite{DiDonato:1982aa}
\begin{equation}
    S_{a,n}(x) = \int_{a}^{x} dt \, P_n(t) \, ,
\end{equation}
\begin{align}
  S_{a,0}(x) & = x - a \, ,\\
  S_{a,n}(x) & = \frac{1}{2n + 1} \left[ P_{n+1}(t) - P_{n-1}(t) \right]^{x}_{a} \, .
\end{align}
for $a = \pm 1$ we get
\begin{align}
  S_{\pm 1, 0}(x) & = x \mp 1 = P_1(x) \mp P_0(x) \, ,\\
  S_{\pm 1, n}(x) & = \frac{1}{2n + 1} \left[ P_{n+1}(x) - P_{n-1}(x) \right] \, ,
\end{align}
where we have used $P_n(\pm 1) = (\pm 1)^n$ to cancel the constant terms.

Returning to the convolution matrices we have for $B^<_{k,n}$ and $n=0$, using $P_0(x) = 1$,
\begin{equation}
\begin{aligned}
  B^\lessgtr_{k, 0} &= 
  \pm
  \frac{2k + 1}{2}
  \sum_{m=0}^M \alpha_m
  \int_{-1}^1 dx \, P_k(x)
  \int_{{\mp 1}}^{x} dt \, P_m(t)
  \\ &= 
  {\pm}
  \frac{2k + 1}{2}
  \sum_{m=0}^M \alpha_m
  \int_{-1}^1 dx \, P_k(x) S_{\mp 1, m}(x)
  \\&=
  {\pm}
  \frac{2k + 1}{2}
  \sum_{m=0}^M  \frac{\alpha_m}{2m+1} 
  \\ \times
  &\int_{-1}^1 dx \, P_k(x) 
  \left[ P_{m+1}(x) - P_{m-1}(x) \right]
  \label{eq:Bk0}
\end{aligned}
\end{equation}
repeatedly using the Legendre orthogonality relation [Eq.\ (\ref{eq:LegOverlap})] gives
\begin{equation}
  B^\lessgtr_{k, 0} =
  \left\{ \begin{array}{lr}
    \alpha_0 {\mp} \frac{\alpha_1}{3} \, , & k = 0 \\[2mm]
    {\pm} 
    (\frac{\alpha_{k-1}}{2k - 1} - \frac{\alpha_{k+1}}{2k + 3}) \, , & \quad k \geq 1
  \end{array}
  \right.
\end{equation}

For the second column with $n=1$ we detail the derivation of $B^<_{k,1}$, the other case $B^>_{k, 1}$ can be done analogously. Using $P_1(x) = x$ we get
\begin{equation}
\begin{aligned}
  B^<_{k, 1} &=
  \frac{2k + 1}{2}
  \sum_{m=0}^M \alpha_m
  \\ &\times
  \int_{-2}^0 \!\!\! dx \, P_k(x + 1)
  \int_{-1}^{x+1} \!\!\!\!\! dt \, P_m(t) P_1(x - t)
  \\ &=
  \frac{2k + 1}{2}
  \sum_{m=0}^M \alpha_m
  \\ &\times
  \int_{-1}^1 dx \, P_k(x)
  \int_{-1}^{x} dt \, P_m(t) (x - t - 1) 
  \\ &=
  - B^<_{k, 0} \\
  &+
  \frac{2k + 1}{2}
  \sum_{m=0}^M \alpha_m
  \int_{-1}^1 dx \, P_k(x)
  \int_{-1}^{x} dt \, P_m(t) \int_{t}^x ds
  \\ &=
  - B^<_{k, 0} \\
  &+
  \frac{2k + 1}{2}
  \sum_{m=0}^M \alpha_m
  \int_{-1}^1 dx \, P_k(x)
 \int_{-1}^x ds \,
  \int_{-1}^{s} dt \, P_m(t)
  \label{eq:Bk1}
\end{aligned}
\end{equation}
where the last step is obtained by changing the order of integration. The last integral relation is a double Volterra integral and can hence be written using $S_{-1,m}(x)$ as
\begin{equation}
\begin{aligned}
  B^<_{k, 1} &=
  - B^<_{k, 0}
  \\
  &+
  \frac{2k + 1}{2}
  \sum_{m=0}^M \alpha_m
  \int_{-1}^1 dx \, P_k(x)
 \int_{-1}^x ds \, S_{-1,m}(s)
  \\ &=
  - B^<_{k, 0}
  \\
  &+
  \frac{1}{2}
  \sum_{m=0}^M \alpha_m
  \int_{-1}^1 dx \,
  [ P_{k-1}(x) - P_{k+1}(x) ]
  S_{-1,m}(x)
\end{aligned}
\end{equation}
where we in the second step have used partial integration and the Legendre derivative relation, Eq.\ (\ref{eq:LegDer}).

For the second case $B^>_{k, 1}$, the only difference is when we change the integration variable, we get $(x-t+1)$ instead of $(x-t-1)$ in Eq.~(\ref{eq:Bk1}), so the sign before $B_{k,0}$ is changed to $+1$.
By using Eq.~(\ref{eq:Bk0}) we obtain the recursion relation
\begin{equation}
  B^\lessgtr_{k, 1} = \mp B^\lessgtr_{k, 0} + \frac{B^\lessgtr_{k-1, 0}}{2k - 1} - \frac{B^\lessgtr_{k+1, 0}}{2k + 3}
  \, , \quad k \ge 1
\end{equation}
with the special case for $k = 0$, $B^\lessgtr_{0, 1} = \mp B^\lessgtr_{1, 0} / 3$.

\clearpage

%

\end{document}